\documentclass[aps,prl,reprint,superscriptaddress,amsmath,amssymb,showpacs,floatfix,longbibliography]{revtex4-1}
\usepackage{graphicx}
\usepackage{bm}
\usepackage{color}
\usepackage{lipsum}

\newcommand{\RN}[1]{\text{\uppercase\expandafter{\romannumeral#1}}}
\newcommand{\bmf}{{\bm{f}}}
\newcommand{\bmx}{{\bm{x}}}

\newcommand{\bmv}{{\bm{v}}}
\newcommand{\bmxi}{{\bm{\xi}}}
\newcommand{\bmB}{{\bm{B}}}
\newcommand{\bmJ}{{\bm{J}}}
\newcommand{\bmeta}{{\bm{\eta}}}
\newcommand{\sfG}{{\mathsf{G}}}
\newcommand{\sfC}{{\mathsf{C}}}
\newcommand{\sfK}{{\mathsf{K}}}
\newcommand{\sfA}{{\mathsf{A}}}
\newcommand{\bmOmega}{{\bm{\Omega}}}

\newcommand{\bmzeta}{{\bm{\zeta}}}

\newcommand{\fm}{{\text{fm}}}
\newcommand{\zm}{{\text{zm}}}
\newcommand{\lm}{{\text{lm}}}

\begin{document}

\title{Emergence of Nonwhite Noise in Langevin Dynamics 
with Magnetic Lorentz Force}

\author{Hyun-Myung Chun}
\affiliation{Department of Physics, University of Seoul, Seoul 02504,
Korea}
\author{Xavier Durang}
\affiliation{Department of Physics, University of Seoul, Seoul 02504,
Korea}
\author{Jae Dong Noh}
\affiliation{Department of Physics, University of Seoul, Seoul 02504,
Korea}
\affiliation{School of Physics, Korea Institute for Advanced Study,
Seoul 02455, Korea}

\date{\today}

\begin{abstract}
We investigate the low mass limit of Langevin dynamics for a charged Brownian
particle driven by the magnetic Lorentz force.
In the low mass limit, velocity variables relaxing quickly are 
coarse-grained out to yield effective
dynamics for position variables.
Without Lorentz force, the low mass limit is equivalent to the high friction
limit. Both cases share the same Langevin equation that is obtained 
by setting the mass to zero in the original Langevin equation. 
The equivalence breaks down in the presence of the Lorentz force. 
The low mass limit turns out to be singular.  The system in
the low mass limit is different from the system with zero mass.
The low mass limit is also different from the large friction limit. 
We derive the effective equations of motion in the low mass limit. 
The resulting stochastic differential equation involves a nonwhite noise 
whose correlation matrix has antisymmetric components.
We demonstrate the importance of the nonwhite noise 
by investigating the heat dissipation by the Brownian particle.

\end{abstract}
\maketitle

Recently, dynamics of Brownian particles driven by velocity-dependent forces
attracts growing interests.
The magnetic Lorentz force is the representative example of the
velocity-dependent
force~\cite{{Sabass:2017hy},{JimenezAquino:2008hz},{JimenezAquino:2011dj},{Kursunoglu:1962eb},{Baura:2013bg},{Czopnik:2001kc},{Kaplan:2009ir},{Kumar:2009db},{Pradhan:2010ir},Jayannavar:2007ed,JimenezAquino:2010jg,
Chatterjee:2011uf,Chen:2014ea}. It can be realized in experimental systems.
In the superionic conductor, e.g., AgI, Ag${}^+$ ions diffuse over the
${\mbox{I}}^-$ ions background.
The Lorentz force can be applied to the $\mbox{Ag}^{+}$ ions with the
external magnetic field~\cite{{Gagliardi:2016ft},{Bonella:2017fs}}.
The active matters are modeled with
velocity-dependent forces. Such phenomenological forces
are adopted in order to study collective phenomena of active
matters~\cite{Romanczuk:2010hn,Romanczuk:2012wn,Grossmann:2012hj,
Marchetti:2013bp,Sevilla:2014ve,Dossetti:2015vk,Shim:2016dp}.
In stochastic thermodynamics, theoretical works focus on the extension of
the entropy production, fluctuation theorems, fluctuation-dissipation
relations, and the detailed balance to thermal systems driven by
velocity-dependent
forces~\cite{Ganguly:2013vk,Chaudhuri:2014vf,Kwon:2016kc,Chaudhuri:2016ke,
Lee:2017ki}.
In this paper, we investigate the low mass limit and the large friction
limit of Langevin dynamics for a charged Brownian particle 
under the uniform external magnetic field.
The magnetic Lorentz force is one of the fundamental forces. 
We will show that a nonwhite noise emerges in the low mass limit  
in the presence of the magnetic Lorentz force.

Without velocity-dependent forces, dynamics of a Brownian particle is 
described by the Langevin equation for its position $\bmx$ and velocity $\bmv$:
\begin{equation}\label{Langevin_eq}
\begin{aligned}
\dot\bmx(t) & = \bmv(t) , \\
m \dot \bmv(t) & = \bmf(\bmx(t)) - \gamma \bmv(t) + \bmxi(t) ,
\end{aligned}
\end{equation}
where $\bmf(\bmx)$ is an external force, $\gamma$ is a friction
coefficient, and $\bmxi(t)$ is the Gaussian white noise satisfying
$\langle \xi_i(t)\rangle = 0$ and $\langle \xi_i(t)\xi_j(s)\rangle = 2
\gamma T \delta_{ij} \delta(t-s)$ with the temperature $T$ of the
environment. We set the Boltzmann constant to unity.
Since the last century, the Langevin equation has served as a
framework for the study of equilibrium and nonequilibrium dynamics of
thermal
systems~\cite{Zwanzig:2001vd,Risken:1996vl,Gardiner:2010tp,Kampen:2011vs}.
It also plays a crucial role in the recent development of 
stochastic thermodynamics, the statistical physics theory at the level of
microscopic stochastic trajectories~\cite{Kurchan:1998us,Seifert:2005vb}.

In experimental situations, the damping force usually dominates the
other forces~\cite{Dhont:1996ub,Gavrilov:2017fn}. 
Then, the velocity relaxes quickly in a time scale $\tau_r=m/\gamma$,
and the inertia term $m \dot\bmv$ becomes negligible for $t\gg \tau_r$.
The effective equations of motion in the limit
are obtained in the following way: 
(i) One considers the Fokker-Planck~(FP) equation for the
probability distribution $P_t(\bmx,\bmv)$ corresponding to the Langevin
equation \eqref{Langevin_eq}. (ii) One then performs the $1/\gamma$
expansion to derive the effective FP equation for the 
coarse-grained probability distribution
$Q_t(\bmx) \equiv \int d\bmv P_t (\bmx,\bmv)$.  The expansion can be done
systematically by using the Brinkman
expansion~\cite{Brinkman:1956bm,Risken:1996vl} or the projection operator
method~\cite{Gardiner:2010tp}. (iii) The FP equation
is transformed back to the Langevin equation. 
The resulting overdamped Langevin equation reads 
\begin{equation}\label{OverdampedLangevin_eq}
\gamma \dot\bmx(t) = \bmf(\bmx(t)) + \bmxi(t) .
\end{equation}
It has the same form as that obtained 
by setting $m=0$ in \eqref{Langevin_eq}. 
Namely, the systems in the large friction limit, in the low mass limit, and
with zero mass are equivalent to each other. They all share the same Langevin
equation~\eqref{OverdampedLangevin_eq}.  
The overdamped limit of the Langevin
equation with multiplicative noises was also
studied~~\cite{Sancho:1982ex,WIDDER:1989go,Sekimoto:1999bk,Matsuo:2000cp,
{Ao:2007ef},Yang:2013jy,Durang:2015gq}.

The large friction limit and the low mass limit of Langevin dynamics has not
been studied thoroughly in the presence of velocity-dependent forces.
Some literatures study the large friction dynamics of Lorentz force systems
by setting $m$ to 
zero~\cite{Jayannavar:2007ed,JimenezAquino:2008hz,JimenezAquino:2010jg,
Chatterjee:2011uf,Chen:2014ea}.
We raise the question whether the equivalence between the low mass limit,
the large friction limit, and the zero mass case is still valid in the
presence of the magnetic Lorentz force,
one of the simplest examples of velocity-dependent forces.
We will derive the stochastic differential equation 
for the motion in the low mass limit. 
It turns out that the low mass limit is singular. Dynamics in the low mass 
limit is different from that with zero mass and from that in the large 
friction  limit.
We discover that a nonwhite noise emerges in the low mass limit. The
nonwhite noise has an intriguing correlation property which has not been 
studied before.  Our work will open up an unexplored avenue  in the study of
stochastic differential equations. It may also 
have an impact on experimental systems such as the 
superionic conduction mentioned earlier.

Suppose that the magnetic field is directed to the $z$ direction,
$\bmB = B_0 \hat{\bm{z}}$. 
The Lorentz force does not have the $z$ component.
Thus, we focus on the two dimensional motion of the Brownian particle. 
The position and the velocity are denoted by the column
vectors $\bmx=(x_1,x_2)^T$ and $\bmv=(v_1,v_2)^T$, where the superscript 
${}^T$ stands for the transpose. 
The Langevin equation becomes $\dot\bmx=\bmv$ and
\begin{equation}\label{Langevin_Lorentz}
m \dot\bmv(t) = \bmf(\bmx(t)) - \mathsf{G} \bmv (t) + \bmxi(t) ,
\end{equation}
where the $2\times 2$ matrix $\mathsf{G}$ is defined as
\begin{equation}\label{g_matrix}
\mathsf{G}=
\begin{pmatrix}
\gamma && -B \\
B && \gamma
\end{pmatrix}
\end{equation}
with $B=qB_0$. 
The external force $\bmf(\bmx) = (f_1(\bmx),f_2(\bmx))^T$ and
the white noise $\bmxi(t) = (\xi_1(t),\xi_2(t))^T$ are also denoted by the 
two dimensional column vectors. 

The equations of motion in the low mass limit may be obtained indirectly 
by using the FP equation representation. 
This method works well for systems 
without velocity-dependent force~\cite{Risken:1996vl}.
The probability distribution $P_t(\bmv,\bmx)$ satisfies 
the FP or Kramer equation 
        \begin{equation} \label{Kramer}
        \partial_t P_t({\bm x},{\bm v}) = \left(L_{rev}+L_{irr}\right)
P_t({\bm x},{\bm v}) ~,
        \end{equation}
where $L_{rev}~(L_{irr})$ is the reversible~(irreversible) part of the 
time evolution operator. They are given by
        \begin{eqnarray}\label{rev.irr}\nonumber
        L_{rev} &=& -\bmv \cdot \bm{\nabla}_{\bm{x}} -
\frac{\bmf}{m}\cdot\bm{\nabla}_\bmv  - 
\frac{q}{m}\bm{\nabla}_\bmv\cdot(\bmv\times \bmB)\\
        L_{irr} &=& \frac{\gamma}{m} \bm{\nabla}_\bmv \cdot \left( \bmv+
\frac{T}{m} \bm{\nabla}_\bmv\right)~.
        \end{eqnarray}
We use the shorthand notation $\partial_\alpha$ for the partial derivative with
respective to a variable $\alpha$. When one takes the cross product, a
two-dimensional vector should be regarded as a three-dimensional one with 
null $z$ component.

Following the standard procedure~\cite{Risken:1996vl}, 
we first rewrite (\ref{Kramer}) in terms of 
$\bar{P}_t(\bmx,\bmv) = [\psi_0(v_1)\psi_0(v_2)]^{-1} 
P_t(\bmx,\bmv)$ and 
$\bar{L}_{rev,irr} = [\psi_0(v_1)\psi_0(v_2)]^{-1} L_{rev,irr}
[\psi_0(v_1)\psi_0(v_2)]^{1}$ with 
$\psi_0(v) \equiv (2\pi T/m)^{-1/4} e^{-mv^2 /(4T)}$.
Then, the transformed distribution is expanded as
\begin{eqnarray}\label{expand}
        \bar{P}_t(\bmx,\bmv) = \sum_{n_1,n_2=0}^{\infty} c_{n_1,n_2}(\bmx,t)
\psi_{n_1}(v_1)\psi_{n_2}(v_2)
\end{eqnarray}
in terms of the orthonormal basis functions 
$\psi_n(v)  \equiv \left( -\sqrt{\frac{T}{m}} \partial_{v} + 
\frac{1}{2}\sqrt{\frac{m}{T}} v_i\right)^n \psi_0(v) / \sqrt{n!}$.
The FP equation yields the coupled differential equations for the 
coefficients $\{c_{n_1,n_2}\}$, called the Brinkman's 
hierarchy~\cite{Risken:1996vl}.
Among all the coefficients, $c_{0,0}(\bmx,t)$ is the most important
one since it is equal to the marginal distribution $Q_t(\bmx)=\int d\bmv
P_t(\bmx,\bmv)$. Orthonormality of $\{\psi_n(v)\}$ ensures the equality
$c_{0,0}(\bmx,t) = Q_t(\bmx)$.
In the low mass limit, the hierarchy is closed within the set of three
coefficients $\{c_{0,0},c_{1,0},c_{0,1}\}$. Introducing the notation
$\bm{c} = (c_{1,0},c_{0,1})^T$, it becomes
$\partial_t c_{0,0} = -\bm\nabla_\bmx\cdot \left(\sqrt{\frac{T}{m}} 
\bm{c}\right)$ and 
$\sqrt{\frac{T}{m}}\bm{c} = \sfG^{-1} (\bmf - T \bm\nabla_\bmx) c_{0,0} +
O(m)$. Combining these equations, we obtain the effective FP equation 
\begin{equation}\label{eq:FPEqOv}
\partial_t Q_t (\bm{x}) = -\bm{\nabla}_{\bm{x}}\cdot \bmJ
\end{equation}
with the probability current
\begin{equation}\label{prob_current}
        \bmJ = \left[\mathsf{G}^{-1} \bm{f}(\bm{x}) - T \mathsf{G}^{-1}
\bm{\nabla}_{\bm{x}}\right] Q_t(\bm{x}) \ .
\end{equation}
The first term represents the drift current and the second term the
diffusion current.
Details of the derivation are presented elsewhere~\cite{unpub}.

The diffusion current has an abnormal form. 
For the Langevin system, the diffusion current is given by the product 
of a {\em symmetric} diffusion matrix and the gradient of the probability 
distribution~\cite{Risken:1996vl,Gardiner:2010tp,Kampen:2011vs}.
By contrast, the matrix $\sfG^{-1}$ has {\em antisymmetric} components 
$(\sfG^{-1})_{12} = - (\sfG^{-1})_{21}$. Such a
diffusion current cannot be realized by any Langevin system. 
As a remedy, one may replace the probability
current $\bmJ$ with $\bmJ_s = \mathsf{G}^{-1} \bmf - T \mathsf{G}_s^{-1}
\bm{\nabla}_\bmx Q$ using the symmetrized matrix $\mathsf{G}_s^{-1} \equiv 
[\mathsf{G}^{-1}+(\mathsf{G}^{-1})^T]/2$.
Noting that $\bm\nabla_\bmx\cdot \sfG^{-1} \bm\nabla_\bmx = \bm\nabla_\bmx
\cdot\sfG_s^{-1} \bm\nabla_\bmx$, one finds that the symmetrized current 
leaves the FP equation \eqref{eq:FPEqOv} unchanged. The symmetrized FP
equation is equivalent to the Langevin equation
\begin{equation}\label{NaiveLimit}
\dot\bmx(t) = \mathsf{G}^{-1} \bmf(\bmx(t)) + \bm{\zeta}(t)
\end{equation}
where $\bmzeta(t)$ is the white noise satisfying
$\langle\bmzeta(t)\rangle=0$ and 
$\langle \bmzeta(t)\bmzeta(s)^T\rangle = 2 T \sfG_s^{-1} \delta(t-s)$.

We notice the equality $\sfG_s^{-1} = \gamma
\sfG^{-1} (\sfG^{-1})^T$ for the specific matrix $\sfG$ in \eqref{g_matrix}.
It implies that the noise $\bmzeta(t)$ has the same statistical property as
$\sfG^{-1}\bmxi(t)$ with the white noise $\bmxi(t)$ in
\eqref{Langevin_Lorentz}. 
Thus, the effective Langevin equation~\eqref{NaiveLimit} is equivalent to 
the one obtained by setting $m$ to zero from the original
Langevin equation~\eqref{Langevin_Lorentz}.
One may be tempted to conclude that 
the low mass limit is also equivalent to the mass zero system in the 
presence of the Lorentz force. However, the Langevin equation~(\ref{NaiveLimit})
does not reproduce the probability current \eqref{prob_current}. 
Furthermore, as will be shown later, the dissipations in the system 
\eqref{NaiveLimit} and \eqref{Langevin_Lorentz} are different from each other 
in the $m\to 0$ limit.
These observations strongly suggest that the Langevin equation 
in \eqref{NaiveLimit} is not the proper low mass limit. 

As the FP equation approach fails, we derive the low mass limit 
from the equations of motion directly. 
We start with the formal solution 
\begin{equation}
\bmv(t) = 
\frac{1}{m}\int_0^t dt' e^{-\mathsf{G}(t-t')/m} 
\left[ \bmf(\bmx(t')) + \bmxi(t')\right]  
\end{equation}
of the Langevin equation \eqref{Langevin_Lorentz}.
We omitted the transient term $e^{-\frac{\mathsf{G}}{m}
t} \bmv(0)$ because it is negligible for finite $t$ in the small $m$ limit. 
The transient term will always be neglected.
The formal solution leads to the stochastic integro-differential 
equation for $\bmx(t)$:
\begin{equation}\label{formal_eq}
\dot \bmx(t) = \frac{1}{m} \int_0^t dt'\ e^{-\mathsf{G}(t-t')/m}
 \bmf(\bmx(t')) + \bmeta_m(t)  \ ,
\end{equation}
where the noise is given by
\begin{equation}
\bmeta_m(t) = \frac{1}{m} \int_0^t dt'\ e^{-\mathsf{G}(t-t')/m} \bmxi(t') \ .
\end{equation}

We first reveal the statistical property of the noise.
The noise $\bmeta_m$ is Gaussian distributed with $\langle \bmeta_m(t) \rangle =
0$ and $\langle \bmeta_m(t) \bmeta_m(s)^T \rangle = \sfC_m(t,s)$, where the
correlation matrix is given by
\begin{equation}
\begin{aligned}
\mathsf{C}_m(t,s) & = \frac{T}{m}
e^{-\frac{1}{m}\left(\mathsf{G}t+\mathsf{G}^T s\right)
+\frac{1}{m} \left(\mathsf{G}+\mathsf{G}^T \right)\min(t,s)} \\
 &= \left\{ 
  \begin{aligned} 
     \frac{T}{m} e^{-\frac{\sfG}{m} (t-s)} & \mbox{\quad if }\ t \geq s \\ 
     \frac{T}{m} e^{-\frac{\sfG^T}{m} (s-t)} & \mbox{\quad if }\ t < s .
   \end{aligned}
\right. 
\end{aligned}
\end{equation}
As it depends on $(t-s)$, we will use the  notation $\sfC_m(t-s)$ 
for the correlation matrix.  It satisfies $\sfC_m(-u)=\sfC_m(u)^T$. 
The elements are given by
\begin{equation}\label{eq:coloredNoise}
\sfC_m(u) =\frac{T}{m}e^{-\frac{\gamma}{m}|u|}
\begin{pmatrix}
\cos\left(\frac{B}{m}u\right) && \sin\left(\frac{B}{m}u\right) \\
-\sin\left(\frac{B}{m}u\right) && \cos\left(\frac{B}{m}u\right)  
\end{pmatrix} .
\end{equation}
The magnetic field generates oscillating antisymmetric off-diagonal 
components.

The correlation functions oscillate with an amplitude decaying exponentially.
As $m$ decreases, they become singular with diverging oscillation 
frequency~$B/m$, vanishing decay time~$m/\gamma$, and diverging
amplitude~$T/m$. 
In order to extract the
limiting behavior, we consider the integral
$I_\alpha \equiv \frac{1}{m}\int_0^\infty du~u^\alpha e^{-\frac{\gamma+iB}{m}
u}$ for $\alpha \geq 0$. A straightforward algebra yields that
\begin{equation}\label{Ialpha}
I_\alpha = \frac{\Gamma(1+\alpha)}{(\gamma + i B)^{1+\alpha}} m^\alpha
\end{equation} 
with the Gamma function $\Gamma(z) = \int_0^\infty dx~ x^{z-1}e^{-x}$.
In the $m\to 0$ limit, only $I_0 = (\gamma-iB)/(\gamma^2 + B^2)$ converges
to a nonzero value while all $I_{\alpha >0}$ vanishes. 
This property yields that
\begin{equation}\label{kernel}
\begin{split}
\lim_{m\to 0} \int_0^\infty du~ h(u)\sfC_m(u)  & =  h(0)T~ \sfG^{-1} \\
\lim_{m\to 0} \int_{-\infty}^0 du~ h(u)\sfC_m(u) & =  h(0)T~ (\sfG^{-1})^T 
\end{split}
\end{equation}  for any function $h(u)$ having
a nonsingular expansion around $u=0$. The second equality comes from 
the symmetry property $\sfC_m(-u) = \sfC_m(u)^T$. 
We introduce the notations $\delta_{\pm}(u)$ as the variants of the Dirac
$\delta$ function. They are equal to zero for $u\neq 0$ while
$\int_0^\infty du\ \delta_+(u) = \int_{-\infty}^0 du\ \delta_-(u) = 1$ and
$\int_0^\infty du\ \delta_-(u) = \int_{-\infty}^0 du\ \delta_+(u) = 0$.
Then, the correlation matrix in the $m\to 0$ limit is represented as 
\begin{equation}\label{noise_correlator}
\sfC(u) \equiv \lim_{m\to 0} 
 \sfC_m(u) = T \sfG^{-1} \delta_+(u) + T (\sfG^{-1})^T \delta_-(u) .
\end{equation}

We next consider the first term in the right hand side of \eqref{formal_eq}.
When one changes the integration variable from $t'$ to $u=(t-t')$, 
it is written as 
$\frac{1}{T} \int_0^t du\ \sfC_m(u) \bmf(\bmx(t-u))$. 
It converges to $\sfG^{-1} \bmf(\bmx(t))$ from \eqref{kernel}. 
Therefore, we finally obtain the effective equations of motion 
in the low mass limit:
\begin{equation}\label{low_mass_limit}
\dot\bmx(t) = \sfG^{-1} \bmf(\bmx(t)) + \bmeta(t) \ ,
\end{equation}
where the noise $\bmeta(t)$ has the correlation matrix $\sfC$ in 
\eqref{noise_correlator}.
It is a nonwhite noise whose correlation matrix $\sfC$ is nonsymmetric.
The antisymmetric components of $\sfC$ makes \eqref{low_mass_limit} 
different from \eqref{NaiveLimit} fundamentally. 

By analogy with the Wiener process $\bm{W}(t) = \int_0^t dt' \bmxi(t')$ with
a white noise $\bmxi(t)$, one may consider the time-integrated quantity
$\bmOmega(t) = \int_0^t dt' \bmeta(t')$.
It will be called the $\Omega$ process.
The statistical properties of
the $\Omega$ process are summarized as 
\begin{eqnarray}
\langle \bmOmega(t) \bmOmega(s)^T \rangle 
    &=& \frac{2\gamma T}{\gamma^2+B^2} \min(t,s) \mathsf{I} \\
\label{Omega-Omega}
\langle \bmOmega(t) \bmeta(s)^T \rangle &=& \left\{
  \begin{aligned}
     \frac{2\gamma T}{\gamma^2+B^2}\mathsf{I}\ &, \mbox{ if } t > s \\
     T(\sfG^{-1})^T\ &, \mbox{ if } t = s \\
     0             \quad\quad &, \mbox{ otherwise}
  \end{aligned}\right. \label{Omega-eta}
\end{eqnarray}
with the identity matrix $\mathsf{I}$.
These are derived by taking the $m\to 0$ limit of 
the corresponding quantities with $\bmeta_m$.

Because of the nonwhite noise $\bmeta(t)$, the stochastic
equation \eqref{low_mass_limit} does not have the corresponding FP equation. 
Nevertheless, the equation governing the time evolution of the probability
distribution can be derived by using the functional derivative
method~\cite{Hanggi:1978cq,Sancho:1982ex,Hanggi:1995vn,Luczka:2005do}.
The probability distribution is given by 
$Q_t(\bm{x}) = \langle \delta(\bm{x}(t)-\bm{x}) \rangle$
where $\bm{x}(t)$ is a functional of the noise $\{\bmeta(s)|0<s<t\}$ 
and $\langle~\rangle$ denotes the average over the
noise realizations.
The time derivative of $Q_t(\bmx)$ involves ${\partial_t}
\delta(\bmx(t)-\bmx) = \left[ \dot\bmx(t) \cdot \bm{\nabla}_{\bmx(t)}
\right] \delta(\bmx(t)-\bmx) =  - \bm{\nabla}_{\bm{x}} \cdot
[\dot{\bm{x}}(t) \delta(\bm{x}(t)-\bm{x})]$, where the last equality is
obtained by using the property of the $\delta$ function. Thus, the time
evolution of $Q_t(\bmx)$ is governed by $\partial_t Q_t(\bmx) =
-\bm\nabla_\bmx \cdot \bmJ(\bmx,t)$ with $\bmJ(\bmx,t) = \langle \dot\bmx(t)
\delta(\bmx(t) - \bmx) \rangle$. Eliminating
$\dot\bmx(t)$ using \eqref{low_mass_limit}, one obtains 
\begin{equation}\label{current2}
\bm{J}(\bm{x},t) = \mathsf{G}^{-1}\bm{f}(\bm{x}) Q_t(\bm{x})
+ \langle \bm{\eta}(t) \delta(\bm{x}(t)-\bm{x}) \rangle~.
\end{equation}
In order to evaluate $\langle \bm{\eta}(t) \delta(\bm{x}(t)-\bm{x})
\rangle$,
we use the Novikov relation~\cite{Novikov:1965uv}
\begin{equation}
\langle \eta_i (t) F[\bm{\eta}] \rangle
= \sum_{j} \int_0^t ds~ C_{ij}(t-s)
\left\langle \frac{\delta F[\bm{\eta}]}{\delta \eta_j(s)} \right\rangle
\end{equation}
for any functional $F[\bmeta]$ with the noise-noise correlation matrix 
$C_{ij}$. Taking $F[\bmeta] = \delta(\bmx(t)-x)$ and 
noting that $\bmx(t)$ is a functional of $\bmeta$, we have
\begin{equation}\label{eq:Novikov_delta_ftn}
\begin{aligned}
\langle \eta_i(t) \delta(\bm{x}(t)-\bm{x}) \rangle
& = - \sum_{j,k} \int_0^t ds~C_{ij}(t-s) \\
& \times \frac{\partial}{\partial x_k}
\left\langle \frac{\delta x_k(t)}{\delta \eta_j(s)}
\delta(\bm{x}(t)-\bm{x}) \right\rangle.
\end{aligned}
\end{equation}
It is a formidable task to find a closed form expression for the functional 
derivative $\delta x_k(t)/\delta \eta_j(s)$ at arbitrary values of $t$ and
$s$. Fortunately, owing to the property of $\sfC$ in \eqref{noise_correlator},
it suffices to consider the functional derivative at $s=t^{-}$. It is given
by $\lim_{s\to t^-}\delta x_k(t) / \delta \eta_j(s)=\delta_{jk}$.
Consequently, the probability current in \eqref{current2} is the same as
that in \eqref{prob_current}. It confirms that the
stochastic differential equations~\eqref{low_mass_limit} are indeed the 
proper equations of motion in the low mass limit.

We add a remark on the large friction limit. In the large $\gamma$ limit, 
the quantity in \eqref{Ialpha} is given by $I_\alpha = \Gamma(1+\alpha)
m^\alpha/\gamma^{1+\alpha}(1+O(B/\gamma))$. It yields that
$\sfC_m(u) = \frac{T}{\gamma} (\delta_+(u)+\delta_-(u)) \mathsf{I} +
O(\gamma^{-2})$. Thus, in the leading order in $1/\gamma$, the equations 
of motion are given by $\gamma \dot\bmx(t) = \bmf(\bmx(t)) + \bmxi(t)$ with
the white noise $\bmxi(t)$ with the variance $2\gamma T$. 
The Lorentz force contributes as a $O(B/\gamma^2)$ correction, and is
discarded in the leading order.  
It shows that the large friction limit is different from the low mass limit.

We demonstrate the crucial role of the nonwhite noise $\bmeta(t)$ 
with a linear system. 
Consider a two-dimensional motion of a Brownian particle of charge
$q$ in the $xy$ plane. 
It is trapped by a conservative harmonic force $\bmf_c(\bmx) = -k \bmx$ 
and driven by a nonconservative rotating force
$\bmf_{nc}(\bmx) =  \epsilon \bmx \times \hat{\bm z}$. The uniform magnetic
field $\bmB = B_0 \hat{\bm z}$ is applied to the $z$ direction.
The Langevin equation reads $\dot{\bmx}(t) = \bmv$ and 
\begin{equation}\label{linear_langevin}
m \dot \bmv(t) = - \sfK \bmx(t) - \sfG \bmv(t) + \bmxi(t) , 
\end{equation}
where the force matrix $\sfK$ is given by
\begin{equation}\label{f_matrix}
\sfK=
\begin{pmatrix}
k && -\epsilon \\
\epsilon && k
\end{pmatrix}.
\end{equation}
and the matrix $\sfG$ is given in \eqref{g_matrix} with $B=q B_0$.
The nonconservative force performs a work on 
the particle and the injected energy is dissipated into the heat bath as a heat.
The linear system has been studied extensively 
for its nontrivial steady state properties and nonequilibrium 
fluctuation theorems of the work and heat~~\cite{Kwon:2011th,Noh:2012tn,
Park:2016ig,Berut:2016ba,Berut:2016ud}.

We focus on the average power $w=\langle \bmf_{nc} \cdot
\bmv\rangle_s =  \epsilon \langle (\bmx \times \hat{\bm z}) \cdot
\bmv\rangle_s = -\epsilon \langle (x_1 v_2 - x_2 v_1)\rangle_s$
of the work done by the nonconservative force $\bmf_{nc}$ 
in the steady state. $\langle~\rangle_s$ denotes the steady state average.
It is equal to the heat dissipation rate in the steady
state. We calculate the power for three systems: 
$w_{\fm}$ from the original Langevin equation in
\eqref{linear_langevin} with finite $m$, $w_\zm$ from \eqref{NaiveLimit} 
where $m$ is set to zero, 
and $w_{\lm}$ from the low mass limit in \eqref{low_mass_limit}.
Since the equations of motion are linear, the average powers
can be obtained analytically. 
They are given by 
\begin{eqnarray}
w_{\fm} &=& \frac{2\epsilon^2 T}{\gamma k+ \epsilon B -
m\epsilon^2/\gamma}~,
\label{w1} \\
w_\zm &=& \frac{2\epsilon^2 T}{\gamma k + \epsilon B} - \frac{2\epsilon
BT}{\gamma^2+B^2}~, \label{w2} \\
w_{\lm} &=& \frac{2\epsilon^2 T}{\gamma k + \epsilon B}~,
\end{eqnarray}
whose derivations are presented elsewhere~\cite{unpub}.
One notices that the $m\to 0$ limit of $w_{\fm}$ converges to $w_{\lm}$ but
not to $w_\zm$. 
It manifests the singular nature of the low mass limit.

We can pinpoint the origin for the discrepancy between $w_\zm$ and $w_\lm$.
The equations of motion \eqref{low_mass_limit}, $\dot\bmx(t) = -\sfA \bmx +
\bmeta(t)$ with $\sfA \equiv \sfG^{-1} \sfK$, have the formal solution
$\bmx(t) = \int_0^t ds\ e^{-\sfA(t-s)} \bmeta(s)$.
The power involves the correlation matrix 
$\left\langle \bmx(t) \dot{\bmx}(t)^T \right\rangle_s =
-\left\langle \bmx(t) \bmx(t)^T\right\rangle_s \sfA^T + \left\langle \bmx(t)
\bmeta(t)^T\right\rangle_s$. We can use the formal solution to evaluate the
correlation functions in terms of the noise-noise correlation matrix $\sfC$.
Especially, the second term becomes $\langle \bmx(t) \bmeta(t)^T\rangle_s =
\int_0^t ds~ e^{-\sfA(t-s)} \sfC(s-t) = T(\sfG^{-1})^T$ using
\eqref{noise_correlator}.
On the contrary, if one adopts the equations of motion \eqref{NaiveLimit}, 
one obtains $\left\langle \bmx(t) \bmzeta(t)^T\right\rangle_s = 
T(\sfG_s^{-1})^T$ which
misses the antisymmetric component of $\sfG^{-1}$. It makes
$w_\zm$ deviate from $w_\lm = \lim_{m\to 0}w_{\fm}$. 
This example demonstrates the importance of the nonwhite nature of
the stochastic noise $\bmeta$.

In summary, we discover a novel type of stochastic dynamics from 
the low mass limit of Langevin dynamics in the presence of the magnetic 
Lorentz force. One cannot obtain the limiting dynamics by
setting the mass to zero. The low mass limit is also different from the
large friction limit.
The stochastic dynamics in the low mass limit is characterized by the nonwhite
noise whose correlation matrix has antisymmetric components. Importance
of the noise correlation is demonstrated in a linear driven system. The
dissipation is correctly accounted for by the nonsymmetric noise
correlations. Our discovery will be relevant for the study of 
driven charged Brownian particles. The stochastic noise 
$\bmeta$ and the corresponding $\Omega$ process are different from the white
noise and the Wiener process. It will be interesting to study the extent to
which the $\Omega$ process and the Wiener process are different. We hope
that our paper trigger thorough and rigorous study on the property of
the nonwhite noise and the associated stochastic differential equation.

This work was supported by the the National Research
Foundation of Korea (NRF) grant funded by the Korea
government (MSIP) (No. 2016R1A2B2013972). We
thank Prof. Hyunggyu Park, Prof. Chulan Kwon, and Prof. Su-Chan Park for
helpful discussions.

\bibliographystyle{apsrev}
\bibliography{paper}

\end{document}